\documentclass[ preprint,
 amsmath,amssymb,
]{revtex4-1}
\usepackage{xparse}
\usepackage{appendix}
\usepackage{graphicx}
\usepackage{dcolumn}
\usepackage{bm}

 \newcommand{\bray}[1]{\ensuremath{\left( #1 \right|}}
\newcommand{\ket}[1]{\ensuremath{\left| #1 \right>}}
\newcommand{\bra}[1]{\ensuremath{\left< #1 \right|}}
\newcommand{\braket}[2]{\ensuremath{\left< #1 \ \vphantom{#2} \right| 
\left. #2 \vphantom{#1} \right>}}
 
\newcommand{\kety}[1]{\ensuremath{\left| #1 \right)}}

\newcommand{\matrixel}[3]{\ensuremath{\left< #1 \vphantom{#3} \right| #2 
\left| #3 \vphantom{#1} \right>}}

\newcommand{\Tr}{\text{Tr}}
\usepackage{bbold}
\begin{document}

\title{Non-equilibrium aspects of integrable models}
\author{Colin Rylands}
\email{crylands@umd.edu}

\affiliation{Joint Quantum Institute and Condensed Matter Theory Center, Department of Physics,
University of Maryland, College Park, Maryland 20742-4111, U.S.A.}

\author{Natan Andrei}
\email{natan@physics.rutgers.edu}
\affiliation{Department of Physics, Rutgers University,
Piscataway, New Jersey 08854, U.S.A.
}

\email{natan@physics.rutgers.edu}
\date{\today}
\begin{abstract}
Driven by breakthroughs in experimental and theoretical techniques, the study of non-equilibrium quantum physics is a rapidly expanding field with many exciting new developments. 
Amongst the manifold ways the topic can be investigated,  one  dimensional system provide a particularly fine platform.   The trifecta of strongly correlated physics, powerful theoretical techniques and  experimental viability have resulted  in a flurry of research activity over the last decade or so. In this review we explore the non equilibrium aspects  of one dimensional systems which are integrable. Through a number of illustrative examples we discuss  non equilibrium phenomena which arise in such models, the role played by integrability and the consequences these have for more generic systems.
\end{abstract}

\maketitle
\tableofcontents
\section{Introduction}

 Over the past two decades the study of nonequilibrium dynamics of  quantum many body systems
 has moved to the forefront of condensed matter physics. This is driven in large part by the significant advances in experimental techniques regarding the preparation, control and measurement of  adiabatically isolated quantum systems.  The long coherence times of these experimental systems combined with the accuracy to which they can be measured make them ideally suited to study far from equilibrium phenomena. 
 
 Ultracold atomic gas experiments  are a  leading exponent in this regard \cite{IBRMP}. Gases of neutral atoms are held in situ using  laser  beams, a magnetic trap or an atom chip, cooled to nano-Kelvin temperatures and subsequently manipulated through a combination of magnetic fields and laser light.
 The versatility and precision of these experiments allows to  continuously vary numerous system parameters  in real time while simultaneously avoiding undesirable side effects like heating or trap loss. Trapped ion systems and optical cavity experiments likewise provide high precision and versatile  platforms for the study of non equilibrium quantum physics \cite{Schneider}. Also more conventional solid state systems such as quantum dots and mesoscopic superconductors offer  settings that realize various  different nonequilibrium scenarios. These experimental advances have motivated  theoretical physicists to develop new tools, both analytical and computational, to provide  a framework with which to predict and understand the results of these experimental achievements. In turn, these insights  feed back to experiments  to create the vibrant field that exists today. 

Many of the systems so studied are one dimensional. Quantum fluctuations are enhanced at low dimensionality and therefore strong correlations appear naturally, being the rule rather than the exception. Moreover,  powerful theoretical tools are available to study these systems, both numerically and analytically. Among them are the renormalization group \cite{WilsonRMP}, conformal field theory \cite{Conformal} or DMRG \cite{White}. Another  remarkable feature of many of these low dimensional experiments is that they can be accurately  modelled by integrable  Hamiltonians, the focus of this review. The notion of integrability in   quantum systems is trickier to pin down than in their classical counterparts. One expects  however  an integrable Hamiltonian  to exhibit an extensive number of nontrivial conserved quantities \cite{Arnold}, or equivalently,   to possess a complete set of eigenstates that have a particular simple structure characterized by a fixed set of momenta and scattering S-matrices, the Bethe Ansatz \cite{Bethe,Orbach, YangYangXXZ1, YangYangXXZ2, Gaudin, YangPRL67, KorepinBook, Sutherland, SamajBajnok}.  The list of integrable models is long, encompassing lattice and continuum models, relativistic and non relativistic models, classical and quantum models and both local and non local interactions. It includes, among others,  the Lieb-Liniger  \cite{LiebLin1,LiebLin2} and Sine-Gordon/ Massive Thirring models \cite{ZamZam, BT} that describe the physics of cold atom gases \cite{CazalillaRMP, 
XWGuan}, the  Heisenberg spin chains \cite{Bethe, Orbach, Baxter} that describe magnetic chains  and are the natural description of trapped ion systems, the Hubbard model \cite{LiebWu}  describing ultra cold fermions on the lattice, the Gross-Neveu models \cite{AndreiLowen} exhibiting asymptotic freedom and dynamical mass generation,
the Dicke and  Jaynes-Cummings models \cite{Jaynes} that are the starting point for many optical cavity experiments,  the Anderson and Kondo impurity models capture the physics of quantum dots \cite{RMP, TWAKM}   and the BCS model of superconductivity reformulated as the  Richardson model \cite{Richardson, Gaudin}. Many of these examples were proposed to describe experimental systems with
their integrability established long after they were introduced and studied. In addition many non-integrable systems
contain integrable limits. For example the Bose-Hubbard model, a popular model to describe ultra cold Bose gases in optical traps which becomes integrable in the strongly repulsive limit \cite{CazalillaRMP, Dutta_2015}.  The non trivial nature of these limits provides a starting point from which to begin studying the full model. Further, in studying the dynamics of systems which are perturbed away from integrability it has been seen that there exist time scales over which the system behaves  as if it were integrable \cite{Berges, Bertini, LevPRX}.

Once the integrability of a model has been established 
 and  explicit expressions for its eigenstates and eigenvalues are obtained one can proceed  further and develop a  full description of its thermodynamic properties in  equilibrium  by a number of standard techniques \cite{YY,Takahashi}.  No such unified methods exist for the study of out-of-equilibrium behavior of these models but the  many approaches developed thus far, both numerical and analytical \cite{CalabreseCardyJstat, EsslerCaux, DeepakAndrei, JamesReprts, DoyonPRX,deNardisGHD}, have led to much progress in the field, in particular in clarifying the role the non trivial conservation laws play in determining the evolution dynamics of such systems. 

In this brief review we will discuss the non-equilibrium dynamics of integrable quantum systems with a specific focus on models which admit solutions via Bethe Ansatz and on the work of the authors and collaborators on the topic. In the next section we introduce the notion of quench dynamics and give an overview of the distinctive role played by integrability in this setting. Subsequently we provide some illustrative examples of the local non-equilibrium phenomena in the presence of integrability, focusing on the Lieb-Liniger model and the Heisenberg spin chain.  After this we discuss global aspects of integrable quench dynamics using the Lieb-Liniger and the  Sine-Gordon models to compute quantum work distributions emphasizing the interaction effects on them. In the final section we conclude and provide a future outlook for the field.

\section{Quench Dynamics}
A popular way to create and observe  a system out of equilibrium  is  the quantum quench \cite{CalabreseCardy1, CalabreseCardy2, MitraRev, AndreiQuench}. Here a system is prepared in some initial state $\ket{\Psi_i}$, typically an eigenstate of an initial Hamiltonian, $H_i$. It is then allowed to evolve in time using another Hamiltonian, $H$ for which $\ket{\Psi_i}$ is not an eigenstate.  The post quench Hamiltonian, $H$, will be taken to be integrable and in particular one which admits solution via the  Bethe Ansatz.  The difference between $H_i$ and $H$ categorizes the type of quench. Changing a mass parameter, $m_i\to m$ or an interaction strength $c_i\to c$ are known as mass and interaction quenches respectively and along with changing an applied  potential or field, $V_i\to V$ are typical examples of global quenches. These are routinely performed in ultracold atom experiments and as their name and nature suggest have consequences for the whole system. Alternatively a local quench can be performed. This  entails  altering some local system parameter, an  example being a change in the coupling of a quantum dot to external leads or the application of a magnetic field on a local degree of freedom. In contrast to the global quench the effects of this local change propagate outwards through the system during its evolution \footnote{Global changes can also lead to local effects as is the case for edge states in a topological wire}. 

The time scale over which the change is made is an important component. If it is  sufficiently long the system will stay in the state that is adiabatically connected to $\ket{\Psi_i}$  and  remain in quasi equilibrium. It should therefore be short enough for the post quench evolution to be non trivial. We will be concerned with the sudden quench, where  the change in parameters happens instantaneously, namely on a time scale shorter than any  natural one  determined by $H$.  
In a sudden quench the state of the system at time $t$ is,
\begin{eqnarray}\label{Psit}
|\Psi_i(t)\rangle =  e^{-iHt}\;|\Psi_i\rangle=    \sum_n \; \braket{n}{\Psi_i}e^{-i E_n t} \;\ket{ \,n}
\end{eqnarray}
where we expanded the evolved state in terms of the normalized  eigenstates of  $H$ denoted $\ket{n}$ with energy $E_n$. It is evident from this equation why this particular non-equilibrium protocol is so popular. First, we are presented with a simple expression for the time evolution  and second, the overlap $\braket{n}{\Psi_i}$ will generically be non zero for states throughout the spectrum of $H$. Therefore, as the system evolves in time, all (or most) of the Hilbert space is visited not just the low energy sector and so  the system is truly out of equilibrium. It is possible also to consider non-sudden quenches, for example ramping  the parameters from initial to final values over a finite time \cite{DoraHaqueZarand, Smacchia, Sharma}, however sudden quenches are the most tractable from a theoretical point of view and now can be readily implemented in experiments. 

The role of integrability can be discerned already at this stage. As mentioned above, the  integrability of  $H$ means there exists an extensive number of local charges, $\mathcal{Q}_n,~n=0,1,\dots$ that commute with both the Hamiltonian and among themselves. Any measurement of $\mathcal{Q}_n$ is then independent of time,
\begin{eqnarray}\label{Fullcounting}
\matrixel{\Psi_i(t)}{e^{\sum_n\beta_n\mathcal{Q}_n}}{\Psi_i(t)}&=&\matrixel{\Psi_i}{e^{\sum_n\beta_n\mathcal{Q}_n}}{\Psi_i}
\end{eqnarray}
where $\beta_n$ are arbitrary constants. We have written this conservation law in a form that highlights the fact that the full distribution of outcomes for measuring $\mathcal{Q}_n$ are determined solely by the  choice of initial state - not just their average but their full counting statistics. 
This strongly constrains the evolution of the system compared the situation where $H$ is a generic non-integrable Hamiltonian, in which case only the energy and perhaps particle number or momentum are conserved.  

The properties of the time evolved state are typically studied  through the evolution of expectation values of local observables $\mathcal{O}$,
\begin{eqnarray}\label{Operator}
\matrixel{\Psi_i(t)}{\mathcal{O}}{\Psi_i(t)}=\sum_{n,n'}\braket{\Psi_i}{n'}\braket{n}{\Psi_i}e^{-i (E_n-E_{n'})t}\matrixel{n'}{\mathcal{O}}{n}.
\end{eqnarray} 
Here one is mostly concerned  with local operators or products thereof such as correlation functions. Through these one can track the  emergence and evolution of correlations in the system as a function of time and initial state. An interesting topic concerns the formation of order in the post quench system or if indeed one can associate such a concept to a system so far from equilibrium. For example if one quenches from a repulsive fermionic model to an attractive one are Cooper pairs formed, or if one releases bosons from an optical lattice do they form a condensate and if so how does its order parameter behave? It has been shown, for example, that in quenches of the BCS and related models that the order parameter exhibits non trivial  oscillatory behavior depending on the initial state and final Hamiltonian parameters \cite{Yuzbashyan3, Yuzbashyan1, Yuzbashyan2}.

Another fundamental question is whether or not such a system  thermalizes.   Specifically, given a small subsystem  (represented by a local operator $\mathcal{O}$)  is it possible for the remainder of the system to  act as bath and in doing so  allow it to be described using a thermodynamic ensemble? 
In a non integrable system it is believed that such thermalization occurs in the long time limit provided the system is large enough \cite{Deutsch, Srednicki}:  $\lim_{t\to\infty}\matrixel{\Psi_i(t)}{\mathcal{O}}{\Psi_i(t)}=\Tr\left[\mathcal{O}\;e^{-\beta H}\right]$, with the inverse temperature $\beta$
determined by the initial value $\epsilon_i= \langle \Psi_i|H|\Psi_i \rangle =  
\Tr\left[\; H\; e^{-\beta H}\right]$. On the other hand, the abundance of conserved quantities in an integrable system over and above those necessary for a Gibbs ensemble suggests that these should also be included. It was hypothesized \cite{GGE} that integrable models will instead thermalize to a generalized Gibbs ensemble (GGE),
\begin{eqnarray}
\lim_{t\to\infty}\matrixel{\Psi_i(t)}{\mathcal{O}}{\Psi_i(t)}&=&\Tr\left[\mathcal{O}e^{-\sum_n\beta_n\mathcal{Q}_n}\right]
\end{eqnarray}
where the constants $\beta_n$ play the role of generalized inverse temperatures and are fixed by the initial state as in \eqref{Fullcounting}. This behavior has been confirmed through theoretical and experimental work in a number of different scenarios involving various initial states and post quench Hamiltonians \cite{CazalillaJstat, VidmarRigol,  EsslerFagottiJstat, CauxJstat,KonikCaux, GGEexp,NC}.

Beyond these local properties one may ask how the system behaves globally. The central quantity of interest in that case is the Loschmidt echo   $L(t)$ (LE)  \cite{Gorin}, the probability that the system returns to its initial state after time $t$. It is given  as  $L(t)=|\mathcal{G}(t)|^2$ where $\mathcal{G}(t)$, the Loschmidt amplitude (LA),  is the overlap between the initial state with its time evolved self,
\begin{eqnarray}\label{LA}
\mathcal{G}(t)= \langle \Psi_i\,| e^{-iHt}\, |\,\Psi_i\rangle=\sum_n |\braket{n}{\Psi_i}|^2e^{-i E_nt}.
\end{eqnarray}
 Through the LE one seeks to quantify how far the system strays from its initial state, to what degree the information in it has decohered and   over what kind of time scales it may occur. This requires a greater understanding of the behavior of the LA; is it monotonic or perhaps displays some oscillatory character or a dynamical phase transition in time \cite{HeylPolKeh, Heylreview, poz3} and furthermore what role integrability plays in this context. 

A related but less studied quantity is the amount of quantum work done during the quench \cite{Goold}. Quantum work, according to the first law of thermodynamics,  is the difference between the initial and final energies of the quenched system  $dW= dE$ (as the system is adiabatically isolated $dQ=0$). It  requires two measurements for its determination and cannot be represented as an observable. The first measurement yields the initial energy $\epsilon_i$ but as the latter measurement is carried out at time $t$   it may  yield any of the energy eigenvalues $E_n$ of $H$ with amplitude  $\langle n\;|e^{-iHt}|\Psi_i\rangle=  e^{-iE_nt}\;\langle n|\, \Psi_i\rangle$. Quantum work $W$ is thus a random variable  and its probability distribution $\mathcal{P}(W)$  is given by \cite{TakLutHan, Silva08},
\begin{eqnarray}\label{P}
\mathcal{P}(W)=\sum_n\delta(W-(E_n-\epsilon_i)) |\braket{n}{\Psi_i}|^2.
\end{eqnarray}
From here  on  we absorb $\epsilon_i$ into $W$, measuring work relative to the initial energy. Comparing with \eqref{LA} we can see that $\mathcal{P}(W)$ is the  Fourier transform of $\mathcal{G}(t)$,
\begin{eqnarray}\label{LAW}
\mathcal{P}(W)=\int_{-\infty}^\infty\frac{\mathrm{d}t}{2\pi} e^{iWt}\mathcal{G}(t).
\end{eqnarray}
and so a measurement of one quantity gives access to the other. Since within the quench protocol one generally prepares a particular initial state for which the energy is known already, this reduces the determination of $\mathcal{P}(W)$ to repeated measurements of the energy after the quench. Moreover, as the energy is conserved in the post quench evolution these can be performed at any time. Measurement of the work statistics of closed, isolated, quantum systems has been carried out in a number of experimental settings already including quantum dots \cite{Tureci, Tureci2, Goold2} and cold atom gases \cite{QuantumWorkmeter}. The global behavior of the system at all times can then be obtained via the relation to the LA. Aside from studying the Loschmidt amplitude and echo, $\mathcal{P}(W)$ is of great interest itself. The form of the expression \eqref{P} is reminiscent of a spectral function\cite{Mahan} and  shares some of the same properties which are particularly sharp in the presence of integrability \cite{SotGamSil,Palmai, Palmai2, RylandsMTM, PerfettoPiroliGambassi,  RylandsAndreiLLwork}. It exhibits singularities similar to the famous Anderson \cite{AndersonXray} and Mahan \cite{Mahan2} effects from the X-ray edge problem and resonances indicating the presence of bound states in the post quench Hamiltonian. One can study how the appearance and nature of these features is affected by  different choices of $\ket{\Psi_i}$ or $H$. 
An alternative avenue is to take seriously the analogy with a spectral function and use $\mathcal{P}(W)$ to perform spectroscopy on the post quench system. When $H$ is integrable it has a preferred basis in terms of quasi particles which have infinite lifetime, allowing to
 determine how good and over what time scale the description in terms of $H$ is. 
 Lastly, measurement of the work distribution allows one to study the  laws of thermodynamics for isolated quantum systems. In particular work fluctuation theorems and other thermodynamic inequalities can be tested \cite{Jarzynski1, Jarzynski2, Crooks,  HeylKehrein}.


\section{Local non-equilibrium dynamics of some integrable models of cold atoms systems}

We now turn to discuss the quench dynamics of some integrable models used in cold atoms experiments. The first model we consider is in the continuum, the Lieb-Liniger model, the second one, the Heisenberg model, is on the  lattice.

\subsection{The Lieb Liniger model}
The model describes systems of  ultracold  gases of neutral bosonic atoms moving in one dimensional  traps and interacting  with each other via a local density  interaction of strength $c$ which can be repulsive $c>0$ or attractive $c<0$. Aside from being an excellent description of the experimental system, it is one of the simpler Hamiltonians for which there exists an exact solution via Bethe Ansatz.
\begin{eqnarray}
H=-\frac{\hbar^2}{2m}\int\mathrm{d}x\,b^\dag(x)\partial_x^2\,b(x)+c\int\mathrm{d}x\,b^\dag(x)b(x)b^\dag(x)b(x)
\end{eqnarray}
Here $b^\dag(x),~b(x)$  create and annihilate  bosons  of mass $m$.    The exact $N$-particle eigenstate is given by  \cite{LiebLin1,LiebLin2},
\begin{eqnarray}\label{Bethestates}
\ket{\{k_j\}}=\int\mathrm{d}^Nx\,\prod^N_{\substack{i,j=1\\ i<j}}\frac{k_i-k_j-ic\,\text{sgn}(x_i-x_j)}{k_i-k_j-ic}\prod_{l=1}^Ne^{ik_lx_l}b^\dag(x_l)\ket{0}
\end{eqnarray}
where $k_j$ are the single particle momenta and  $\text{sgn}(x)=x/|x|$ is the sign function. The single particle momenta are related to the conserved charges by $q_n=\sum_{j=1}^Nk_j^n$, $q_2$ being proportional to the energy and if we impose periodic boundary conditions, for system size $L$, are quantized according to the Bethe Ansatz equations
\begin{eqnarray}
k_j=\frac{2\pi}{L}n_j-\frac{1}{L}\sum_{l\neq j}\varphi(k_j-k_l)
\end{eqnarray}
Here $\varphi(x)=2\arctan{(x/c)}$ is the two particle phase shift and $n_j$ are distinct integer or half integers for $N$  odd or even respectively. The collection $\{n_j\}$  serve as the quantum numbers of the system and upon adopting the convention $n_j<n_l$ for $j<l$ uniquely identify the states. These form a complete basis and can be used to construct a resolution of the identity,
\begin{eqnarray}\label{Identity}
\mathbb{1}_N=\sum_{n_1<\dots<n_N}\frac{\ket{\{n_j\}}\bra{\{n_j\}}}{\mathcal{N}(\{n_j\})}
\end{eqnarray}
Here we take care not over count the states by ordering the quantum numbers in the sum and have introduced the normalisation of the states which for periodic boundary conditions is \cite{Gaudin, Korepin}
\begin{eqnarray}
\mathcal{N}(\{n\})=\det\left[\delta_{jk}\left(L+\sum_{l=1}^N\varphi'(\lambda_j-\lambda_l)\right)-\varphi'(\lambda_j-\lambda_k)\right].
\end{eqnarray}
Using this identity one can  determine the post quench evolution of the system \`a la eqn \eqref{Psit}. A major stumbling block to proceeding further lies in calculating the overlaps $\braket{\{n_j\}}{\Psi_i}$ which proves to be quite difficult\cite{Brockmann, deNardis}. For a particular class of initial states this task can be greatly simplified by choosing instead an alternate form of the identity. Using properties of the eigenstates  the ordering in momentum space can be exchanged for ordering in coordinate space, a trick first used by Yudson et. al. when studying the Dicke model out of equilibrium  \cite{Yudson}. The alternate form of the identity is \cite{GoldsteinAndrei},
\begin{eqnarray}\label{IdentityY}
\mathbb{1}_N=\sum_{n_1,\dots,n_N}\frac{\ket{\{n_j\}}\bray{\{n_j\}}}{\mathcal{N}(\{n_j\})},
\end{eqnarray}
where we  introduced the notation $\kety{\{n_j\}}$ to denote what we refer to as the Yudson state
\begin{eqnarray}\label{Yud}
\kety{\{n\}}=\int\mathrm{d}^Nx\,\theta(\vec{x})\prod_l^N e^{ik_lx_l}\, b^{\dagger}(x_l)\ket{0}
\end{eqnarray}
with $\theta(\vec{x})$ denoting a Heaviside function which is non zero only for the ordering $x_1>x_2>\dots >x_N$. The Yudson state is  simpler to work with than the full eigenstates of the model and the overlaps with the initial state can be more readily calculated, particularly so if the initial state is similarly ordered in coordinate space.

\subsubsection{ Repulsive interactions}

\begin{figure}

\includegraphics[trim=2cm  2cm 2cm 0cm ,clip=true, width=.65\textwidth]{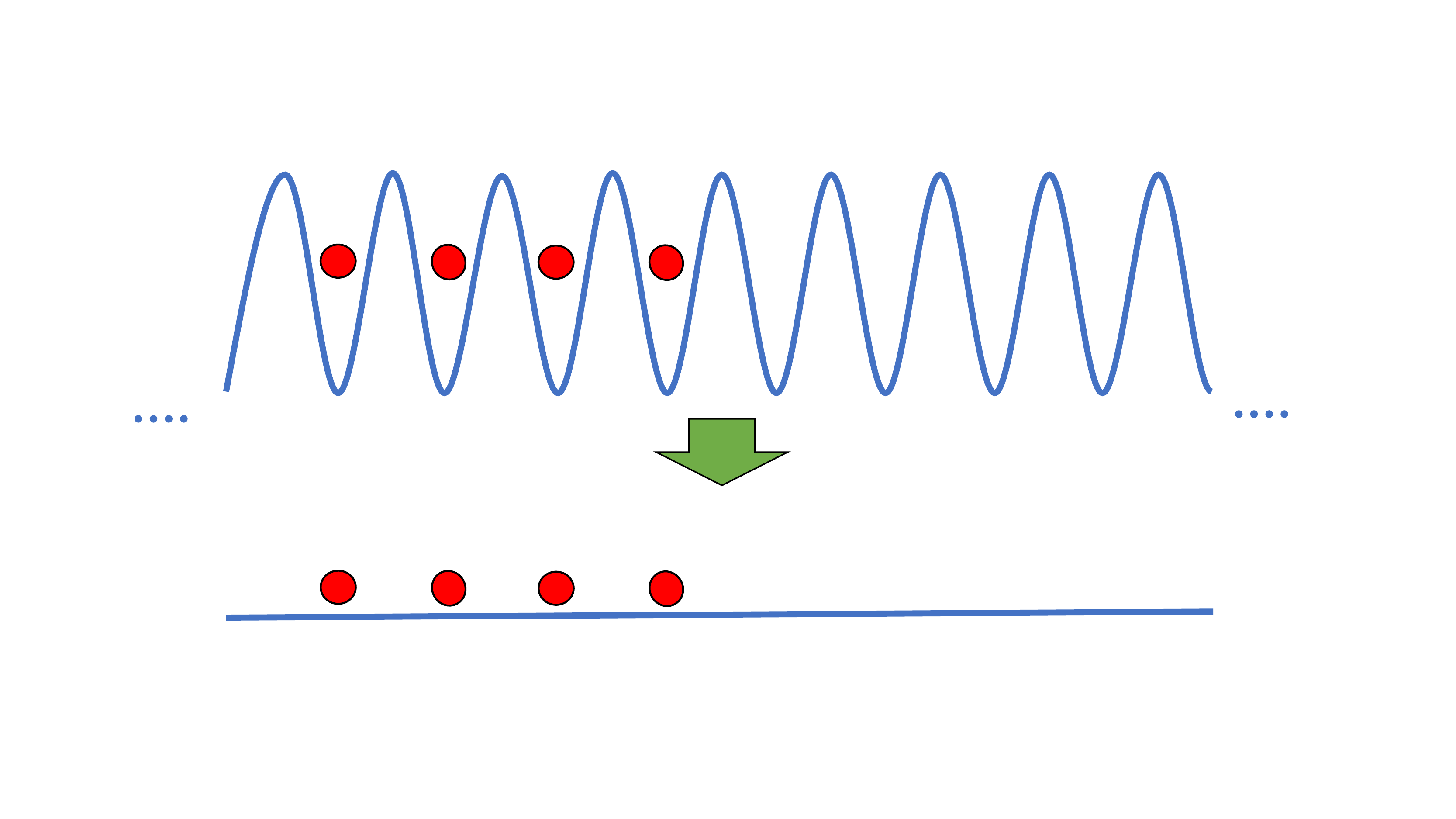}
\caption{The domain wall initial state . A cold atom gas is held in the left part of a deep optical lattice. This is then removed and the gas allowed to expand.  }\label{FigDomain}
\end{figure}
\begin{figure}
 \includegraphics[trim=4cm  9cm 4cm 0cm ,clip=true, width=.9\textwidth]{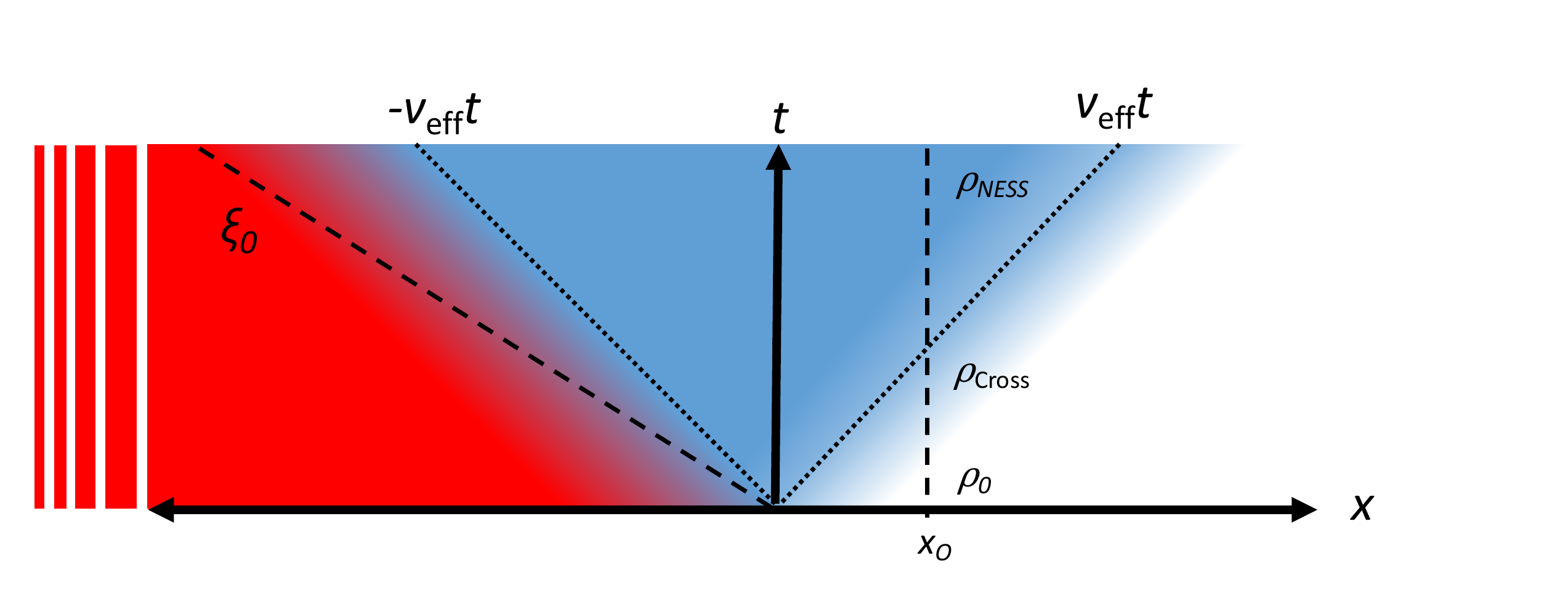}
\caption{ At long times a non-equilibrium steady state is established as depicted on the right. Information about the domain wall propagates away forming several regions similar to a light cone effect. The lack of Lorentz invariance means that this cone is not sharp but rather distinguishes between the NESS, and the asymptotic regions which are unaffected with a broad crossover region separating them. Measuring the density at $x=x_O$  one will see the inital density, $\rho_0$ change to the crossover regime $\rho_\text{Cross}$. The NESS then is established and $\rho_\text{NESS}$ is measured thereafter. }\label{Figlightcone}
\end{figure}

 
Consider the quench where a cold atom gas is initially held in a very deep optical lattice which is then abruptly removed. The initial state is
\begin{eqnarray}\label{PSI0}
\ket{\Psi_i}=\int\mathrm{d}^Nx\prod_{j=1}^N \left[\frac{m\omega}{\pi\hbar}\right]^{\frac{1}{4}}e^{-\frac{m\omega}{2\hbar}(x_j-\bar{x}_j)^2}b^\dag(x_j)\ket{0}
\end{eqnarray}
which describes the ground state of an optical lattice. For simplicity we take only a single boson per site with $\bar{x}_j>\bar{x}_l, ~j<l$ being the position of the initially occupied sites. Utilizing the alternate form of the identity the time evolution is  found to be,
\begin{eqnarray}\label{PSIt}
\ket{\Psi_i(t)}=\left[\frac{4\pi\hbar}{m\omega}\right]^{\frac{N}{4}}\sum_{n_1,\dots,n_N}\frac{e^{-\sum_{j=1}^N\left[\frac{\hbar }{2m\omega}\left(1+i\hbar\omega t\right)k_j^2+ik_j\bar{x}_j\right]}}{\mathcal{N}(\{n_j\})}\ket{\{n_j\}}.
\end{eqnarray}
A particularly enlightening use of the above formula is to study the dynamics of the system in the thermodynamic limit when $\ket{\Psi_i}$ contains a domain wall. This initial state, depicted in FIG. \ref{FigDomain} consists of all lattice sites to the left being filled while those to the right are empty, i.e $\bar{x}_j\leq0$ along with $\bar{x}_{j}-\bar{x}_{j+1}=\delta=L/N,~\forall j$. When the lattice is removed the gas  expands and the particle density will become nonzero between the lattice sites and also to the right of the domain wall. This quench neatly captures aspects of both a local and global quench as follows. 
In the vicinity of the domain wall particles will begin to vacate the left hand side of the system and populate the right hand side, see FIG. \ref{Figlightcone}. Even though the Lieb-Liniger model does not posses Lorentz invariance, the effects of this  quench can only be felt within a light cone centered at the edge. There is no maximum velocity within the model and so in principle excitations can propagate with arbitrary velocity away from the domain wall. The tight initial confinement of the bosons restricts this and the  light cone is determined by a finite effective velocity, $v^\text{eff}$ which depends upon $\omega$. Far to the right and left  the effects of this redistribution of particle density will not be felt, on the right, $x\gg v^\text{eff}t$ the density will remain zero while to the left, $x\ll -v^\text{eff}t$ , the average density will remain $1/\delta$. In the latter region however the effects of the global nature of the quench are still felt as the initially confined bosons will expand and begin to interact with each other. 

We first examine the local portion of the quench. Since to the left there is an infinite particle reservoir and to the right an infinite particle drain the system will never equilibrate, however at long times a non-equilibrium steady state (NESS) consisting of a left to right particle current is established.
This can be investigated by  computing the expectation value  of the density
\begin{eqnarray}
\rho(x,t)=\matrixel{\Psi_i(t)}{b^\dag(x)b(x)}{\Psi_i(t)}
\end{eqnarray}
Utilizing the known formulae for the matrix elements of the density operator with Bethe eigenstates  \cite{Korepin} this can be calculated exactly. To the right of the domain wall, at long times and to leading order in $1/c\delta$ three regions emerge \cite{GoldsteinAndrei}
\begin{eqnarray}
\rho(x,t) =\begin{cases}
\rho_\text{NESS}=\frac{1}{2\delta}-\frac{4\pi}{c\delta^2}
&\frac{m\omega}{2\hbar}\ll x\ll v^\text{eff} t\\
\rho_{\text{Cross}}(x)=\frac{1}{\delta}f+\frac{16}{\pi c\delta^2}\left[e^{-\frac{x^2}{\sigma}}\frac{x\sqrt{\pi}}{\sqrt{\sigma}}f-\frac{1}{2}e^{-2\frac{x^2}{\sigma}}+\frac{\pi}{2}f(1-f)\right]&x\sim v^\text{eff}t\\
\rho_0=0&x\gg v^\text{eff} t
\end{cases}
\end{eqnarray}
where $f=f(x,t)=\frac{1}{2}\text{erfc}\left(\frac{x}{\sqrt{\sigma}}\right)$ and $\sigma=\frac{4\hbar t^2}{m\omega}+\frac{m\omega}{2\hbar}$. Far to the right $x\gg v^\text{eff}t $ we see that the density vanishes while closer to the light cone some complicated crossover behavior occurs. Since the model is Galilean rather than Lorentz invariant the light cone is not sharp giving instead this crossover regime. Most interesting is the region deep inside the light cone in which the density is independent of $x,t$. This signifies the existence of the NESS with the particle density being reduced by the repulsive interactions. This nonequilibrium effect of an open system is to be contrasted with the behavior in a closed system where the system will reach equilibrium with a density $\rho=1/2\delta$.
Within this region all local properties of the system can be calculated by taking the expectation value with respect to this NESS, $\left<\mathcal{O}(x,t)\right>=\matrixel{\Psi_\text{NESS}}{\mathcal{O}}{\Psi_\text{NESS}}$ where $\ket{\Psi_\text{NESS}}$ can be determined by taking the appropriate limit of \eqref{PSIt}.
\begin{figure}
  \includegraphics[width=.65\textwidth]{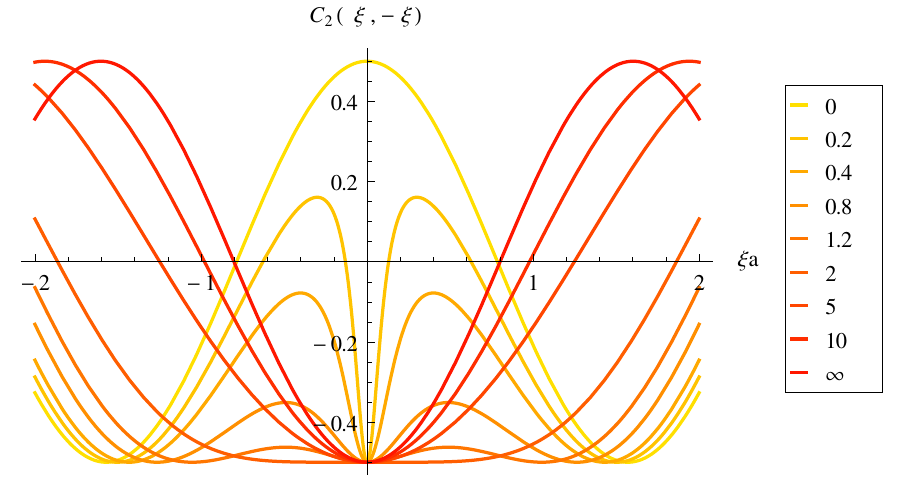}
\caption{ The noise correlation function $C_2(\xi,-\xi)$ , as a function of $\xi=x/\tau$ at long times for a quench from a lattice like initial state. For arbitrary values of $c>0$, with $\delta$ fixed, the system develops a distinct fermionic dip at the origin. Figure taken from \cite{DeepakGuanAndrei} }\label{FigNoise}
\end{figure} 

A mirror of the light cone exists in the left of the system with a NESS region and complicated crossover behavior. 
For $x\ll -v^\text{eff}t$ we are outside the light cone, the system is unaffected by the impurity portion of the quench and the translational invariance is restored. At long times the density within this region is
\begin{eqnarray}
\rho(x,t)=\frac{1}{\delta}\left[1+\sum_{s=1}^\infty e^{-\sigma\frac{\pi^2s^2}{\delta^2}}\cos{\left(\frac{2\pi s x}{\delta}\right)}\right]
\end{eqnarray}
which describes small oscillation about a uniform density of $1/\delta$. Note that it is independent of the interactions and coincides with what one would expect for the Tonks-Girardeau (TG) gas, the $c\to\infty$ limit of the  LL model. To understand this one should go beyond the density and compute the normalised noise correlation function $C_2(x,x')=\frac{\rho_2(x,x',t)}{\rho(x,t)\rho(x',t)}-1$ where
\begin{eqnarray}
\rho_2(x,x',t)=\matrixel{\Psi_i(t)}{b^\dag(x)b(x)b^\dag(x')b(x')}{\Psi_i(t)}.
\end{eqnarray}
As we are interested in the region outside the light cone we shift the origin and measure the coordinates with respect to a point, $x_0\ll -v^\text{eff}\tau$ and $\tau$ is some large time. This correlation function is related to the Hanbury-Brown Twiss  effect and will detect the nature of the interactions between particles, a peak indicating bosons while a dip indicates fermions \cite{HBT, HBT2}. The noise correlation function is computed  by inserting two copies of the identity and evaluating the integrals at long time by saddle point method \cite{DeepakAndrei}. The results for a range of values of $c\delta$ are plotted in FIG. \ref{FigNoise}. It becomes a function only of the ray variables $\xi=x/\tau,~\xi'=x'/\tau$ (measured with respect to $\xi_0=x_0/\tau$ see FIG. \ref{Figlightcone}). For sufficiently long times $\xi\sim 0$ a distinct fermionic dip is seen for arbitrary $c>0$  while $c=0$ shows a bosonic peak, the turn over to the dip occurring on the time scale, $t\sim c^{-2}$. In fact, one can show that the effective coupling constant $c$ increases in time and starting from any initial repulsive value it will flow to strong coupling in the long time limit \cite{Jukic,Jukic2,Jukic3} doing so like $\sqrt{t}$ \cite{DeepakAndrei}. Thus at long times the system  will behave as if it consisted of non interacting fermions. The development of fermionic correlations at large time is known as dynamical fermionization and is the cause of the $c$ independent density outside of the light cone. Subsequently, this was  observed in experiment \cite{DynamicalExp, Weiss}.

\begin{figure}
  \includegraphics[width=.5\textwidth]{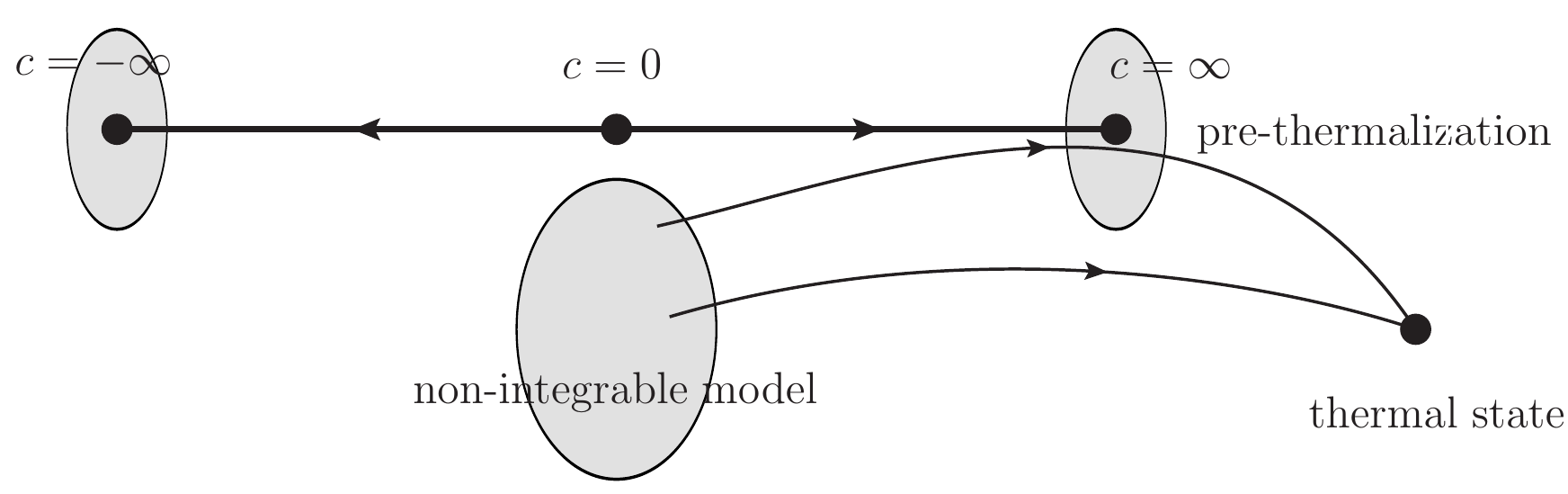}
\caption{ Prethermalization in the Bose Hubbard model, Figure taken from \citenum{DeepakGuanAndrei} }\label{FigDynamicalRG}
\end{figure}

This flow naturally leads to the concept of renormalisation group (RG) flow  in time $t$.  By analogy with conventional  RG ideas, increasing time plays the role of reducing the cut off  with $c=\infty$ being a strong coupling fixed point. For comparison we can recall that in the usual RG picture $c$ has scaling dimension 1 and so also flows to strong coupling. Subsequently, similar behavior was also seen in strongly coupled impurity models \cite{VasseurRL, CrossoverVasseur}. 
Extending the dynamical RG analogy  one can envisage that other Hamiltonians close to the Lieb-Linger  will flow close the neighborhood of the same strong coupling fixed point, prethermalize in other words, only to end up thermalized on  longer time scales if the model is not integrable, see FIG. \ref{FigDynamicalRG}. An example is provided by  the lattice version of the Lieb Liniger model,  the Bose-Hubbard model \cite{DeepakGuanAndrei}. 

\subsubsection{Attractive interactions}
The  attractive regime  is less studied despite the fact that a negative coupling can also be readily achieved experimentally \cite{IBRMP}. Partly the reason for this is the  much richer spectrum for the attractive model which includes bound states of two or more particles which complicates analytic calculations \cite{Maguire, PiroliCalabreseEssler1, PiroliCalabreseEssler2, ZillWright, YurovskyMalomed}. These bound states appear as solutions to the Bethe equations with complex momenta, for example a two particle bound state has Im$(k-k')=c$.
While the  previously quoted identity \eqref{IdentityY} is still formally valid provided the sum also include the bound states, it is no longer practicable as the normalisation is not known in closed form. 

\begin{figure}
 (a)\includegraphics[width=.45\textwidth]{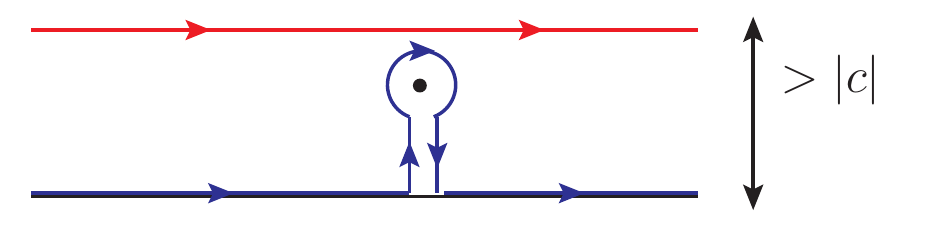}
 (b) \includegraphics[width=.45\textwidth]{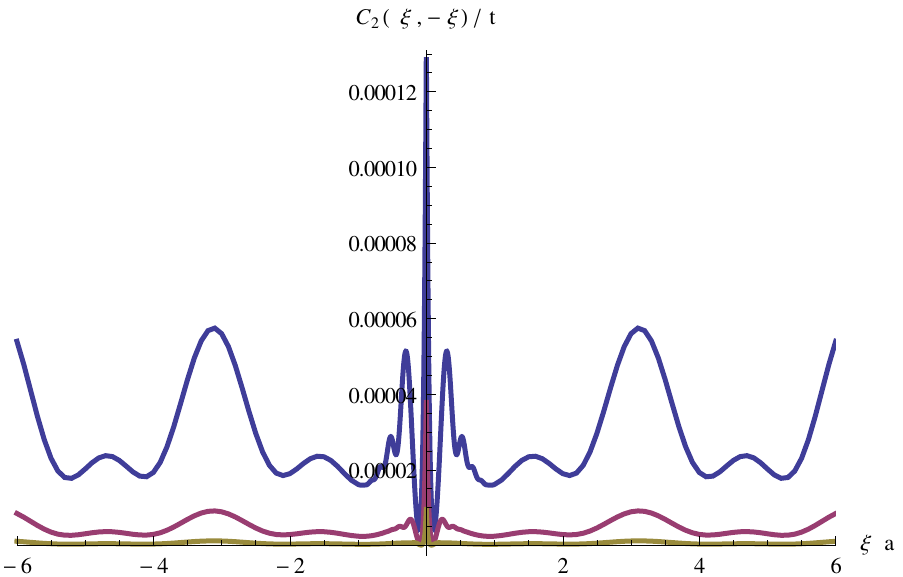}
\caption{(a)The contour integral in \eqref{Identitycont} can be deformed to involve only integrals on the real line. In doing this we pick up the poles contributions as illustrated here. These provide the bound state part of the identity\eqref{Identityatt}. (b) The noise correlation function for attractive bosons in the region outside the lightcone at increasing times, blue being the shortest and yellow the longest. In contrast to the repulsive case we see that a peak at the origin indicative of bound states. The strength of the peak is increasing over time. Figure taken from \cite{DeepakAndrei}.  }\label{FigBound}
\end{figure}

If one works directly on the infinite line however, or equivalently at low density, the identity can be calculated and is given by a simple integral form, \cite{DeepakGuanAndrei}
\begin{eqnarray}\label{Identitycont}
\mathbb{1}=\int_\Gamma\frac{\mathrm{d}^Nk}{(2\pi)^N}\ket{\{k_j\}}\bray{\{k_j\}}
\end{eqnarray}
where the contours of integration $\Gamma=\otimes_{i=1}^N \gamma_i$ are separated out in the imaginary direction
 such that Im$(k_j-k_{j+1})>|c|$ and states are labelled by their single particle momenta $\{k_j\}$ as the $n_j$ are not quantum numbers on the infinite line. This choice of contour is enough to capture all bound states of the model. Via explicit computation using \eqref{Bethestates} it can be confirmed that this is indeed a resolution of the identity and furthermore by deforming the contours to the real line, $\mathbb{R}$ one can write this as 
\begin{eqnarray}\label{Identityatt}
\mathbb{1}=\int_\mathbb{R}\frac{\mathrm{d}^Nk}{(2\pi)^N}\ket{\{k_j\}}\bray{\{k_j\}}+\text{bound states}
\end{eqnarray}
where the bound states come from  the pole contributions at $k_j-k_l=i |c|$ when the contours are deformed to the real line, see FIG. \ref{FigBound}. These bound state contributions coincide exactly with what one would expect from the string hypothesis\cite{Takahashi}.  It is important however  to note that the bound states (the strings) follow directly from the formalism and have not been introduced by hand \cite{DeepakGuanAndrei}. The non bound state part of the identity is formally identical to the that of the repulsive case, only now $c<0$. 

The bound states will enter the quench dynamics in a way that depends on the initial state.
Reconsidering  the domain wall quench for the attractive LL model we need to include  the bound state contributions. However it can shown that the formation of say $n-$particle bound states from an initial state where the bosons are arranged in a lattice  with a constant $\delta$  is exponentially suppressed as $e^{-n|c|\delta}$,  as can be seen upon acting with \eqref{Identitycont} on $\ket{\Psi_i}$ and then deforming the contours. Depending then on the initial state different types of evolution dynamics may ensue. For 
$\delta c \sim 1$ the overlap between the initial state  Gaussians will allow the formation of bound states
and these will dominate the long term evolution of the normalized noise, see FIG.\ref{FigBound}.

On the other hand if the 
the lattice spacing is large enough so that $\delta c \gg 1$ then
to leading order these bound state contributions  can be dropped and we obtain  obtain the same results but now with $c<0$. In particular the NESS density is enhanced by the attractive interactions. In the left half region we similarly get that the noise correlation function is the same but with negative $c$. This simple continuation of the repulsive results to the attractive regime has been encountered before most notably in the super-TG gas\cite{SuperTG1}. The super-TG gas is perhaps one of the earliest examples of an interaction quench \cite{SuperTGExp}, a cold atomic gas is initially prepared in the TG regime and allowed to equilibriate in the trap. The fermionic correlations of the gas cause the bosons to have zero spatial overlap, similar in nature to our lattice like initial state. The interaction strength is then quickly changed from large and positive  to large and negative. Due to the initial profile of the system bound states do not form and a metastable gas is created with many of its properties being obtained via continuation from the repulsive values \cite{SuperTonks2,SuperTG3,SuperTG4}.

\subsection{The XXZ Heisenberg spin chain}

\begin{figure}
  \includegraphics[trim=0 0 11cm 0cm,clip=true, width=.8\textwidth]{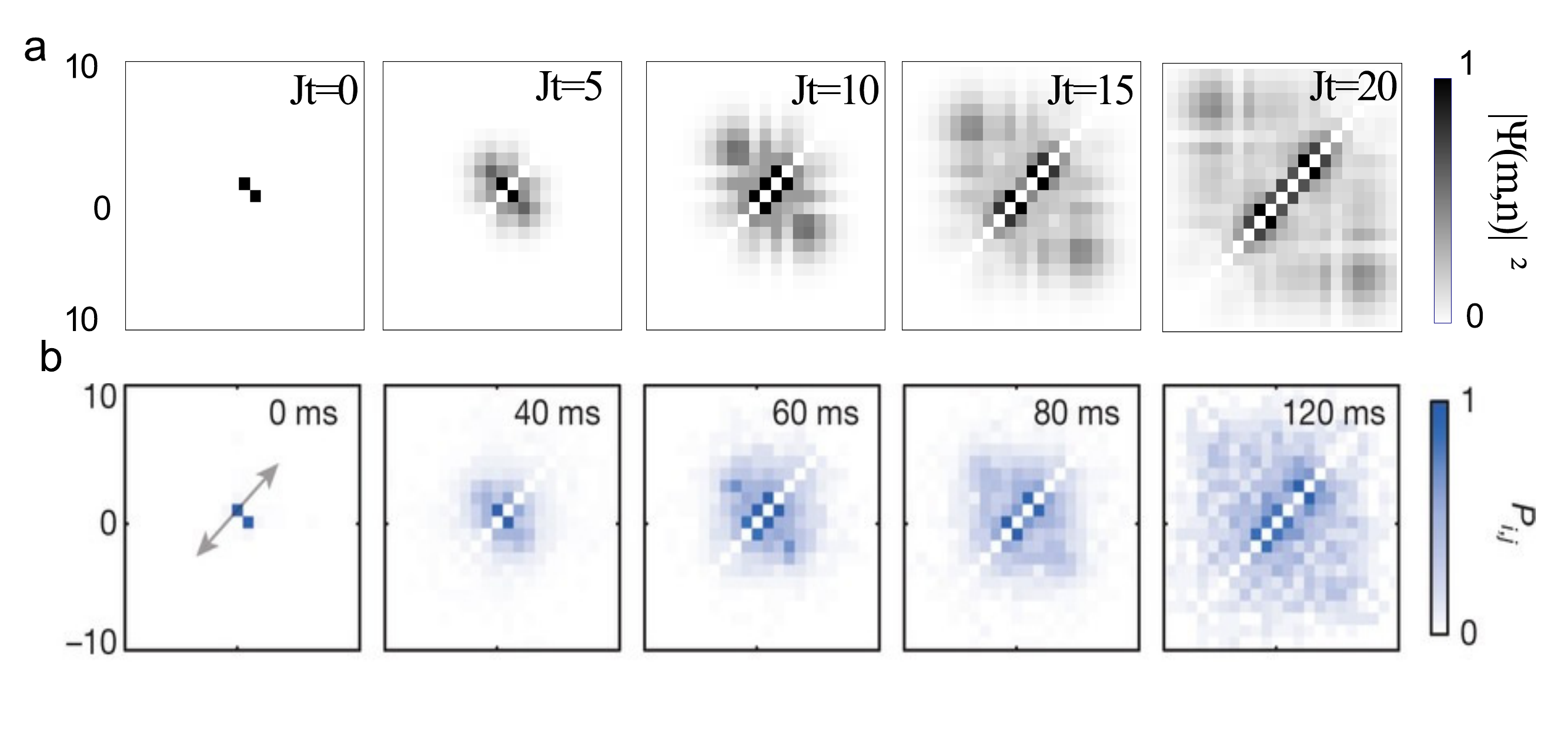}
  \caption{(a) The norm of the wavefunction  $|\Psi(m,n,t)|^2$ at different times for two flipped spins initially at $m=1,n=0$. (b) The joint probabilities at different times of two spins at sites $i$ and $j$ initially at $i=1,j=0$, measured experimentally in \cite{Fukuhara}.
  Figure taken from \cite{LiuAndrei}. }\label{FigXXZ1}
  \end{figure}

\begin{figure}
 \includegraphics[trim=0 0 0 9.05cm,clip=true, width=.8\textwidth]{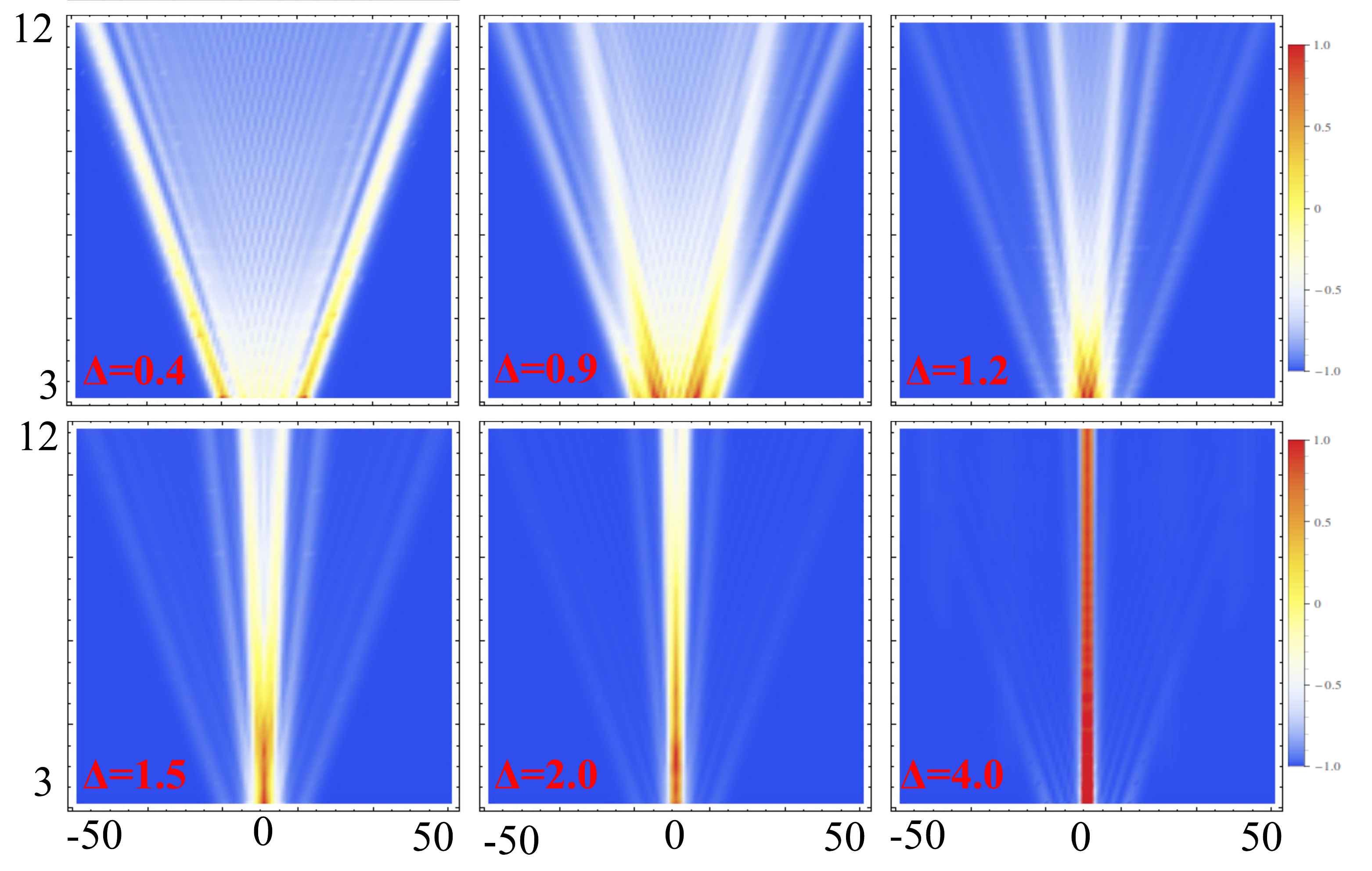}
  \caption{The local magnetization after  a quench from an initial state of 3 flipped spins at the origin for different values of the anisotropy $\Delta$. Time, the vertical direction, is measured in units of the exchange coupling $J$.
  Figure taken from \cite{LiuAndrei}. }\label{FigXXZ}
  \end{figure}

The XXZ Heisenberg chain provides another exemplar of an experimentally relevant integrable model.  The Hamiltonian,
\begin{eqnarray}\label{Heisenberg}
H=J \,\sum_{j=1}^N \{ \sigma^x_j \,\sigma^x_{j+1} +\sigma^y_j\, \sigma^y_{j+1} +\Delta \,\sigma^z_j \,\sigma^z_{j+1} \}
\end{eqnarray}
models a linear array of spin interacting via anisotropic spin exchange. 
The isotropic case $\Delta=1$ is $SU(2)$ invariant and enjoys the distinction of being the first model solved by Bethe  by means of the approach that bears his name \cite{Bethe}. The generalization to the anisotropic case was  given by Orbach \cite{Orbach}. The eigenstates are again characterized by a set of Bethe momenta $\{k_j\}$  describing the motion of $M$ down-spins in a background of $N-M$ up-spins, and are given by:

\begin{eqnarray}
|\vec{k}\rangle = \sum_{\{m_j\}}\prod_{i<j}\,[\theta(m_i-m_j) +s(k_i,k_j)\theta(m_j-m_i)] \prod_j e^{ik_j m_j}\,\sigma^-_{m_j}\,|\Uparrow\rangle
\end{eqnarray}
where $m_j$ the position of the $j$th down spin is summed from 1 to $N$ (the length of the chain), and the S-matrix is given by,
\begin{eqnarray}\label{Heisenberg S matrix}
 s(k_i,k_j)= -\frac{1+e^{i k_i+ ik_j} -2\Delta e^{ik_i}}{1+e^{i k_i+ik_j} -2\Delta e^{ik_j}}
\end{eqnarray}
Similar to the attractive Lieb Linger gas the Heisenberg chain exhibits a complex spectrum which includes bound states in all parameter regimes. To carry out the quench dynamics for the model one needs to construct the appropriate contour  Yudson representation  and use it to time evolve any initial state \cite{LiuAndrei}. Here we display in FIG. \ref{FigXXZ1} the time evolving wave function of two adjacent flipped spins in the background of an infinite number of unflipped spins and compare it to the experimental results (no adjustable parameters are involved.) As in the attractive Lieb-Liniger model the long time limit is dominated by the bound states.  
The time evolution of the magnetization from an initial state of three flipped spins for different values of the anisotropy $\Delta$ is given in FIG. \ref{FigXXZ}.  We see that excitations propagate outward after the quench forming a sharp light cone in contrast to the Lieb-Liniger model. The boundary of the light cone arises from the propagation of free magnons which travel with the maximum  velocity allowed by the lattice. Rays within the lightcone are the propagation of spinon bound states.  As the anisotropy $\Delta$ is increased the bound states slow down and more spectral weight is shifted to them. Due to the integrability of \eqref{Heisenberg} these excitations have infinite lifetime which prevents any dispersion of these features. The introduction of integrability breaking terms can therefore be expected to alter this picture, for example through spinon decay \cite{GrohaSpinon}.

\section{Global non-equilibrium behavior - The Loschmidt Amplitude and Quantum Work}

\begin{figure}
  (a) 
  \includegraphics[width=.45\textwidth]{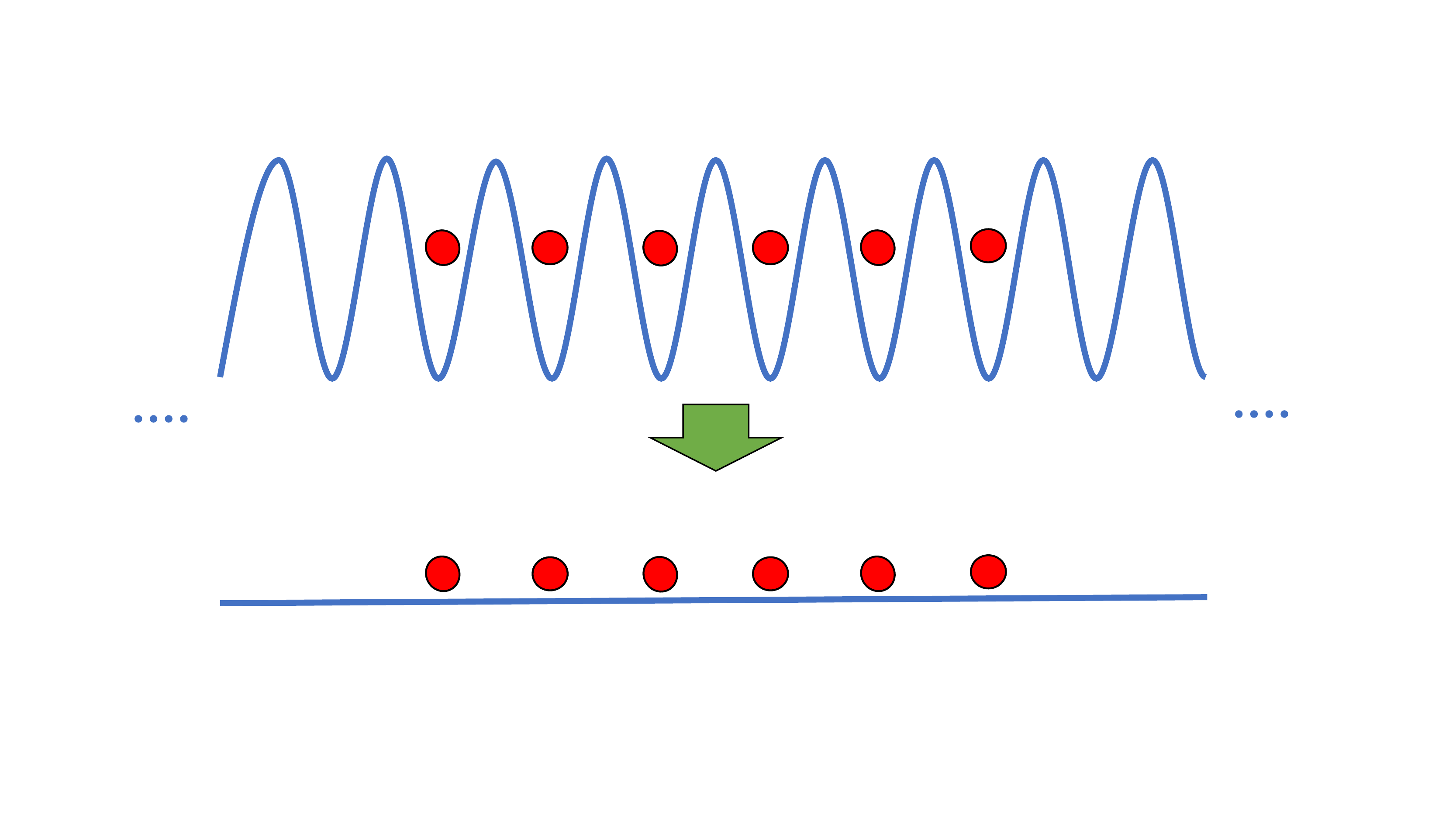}
  (b)
 \includegraphics[width=.45\textwidth]{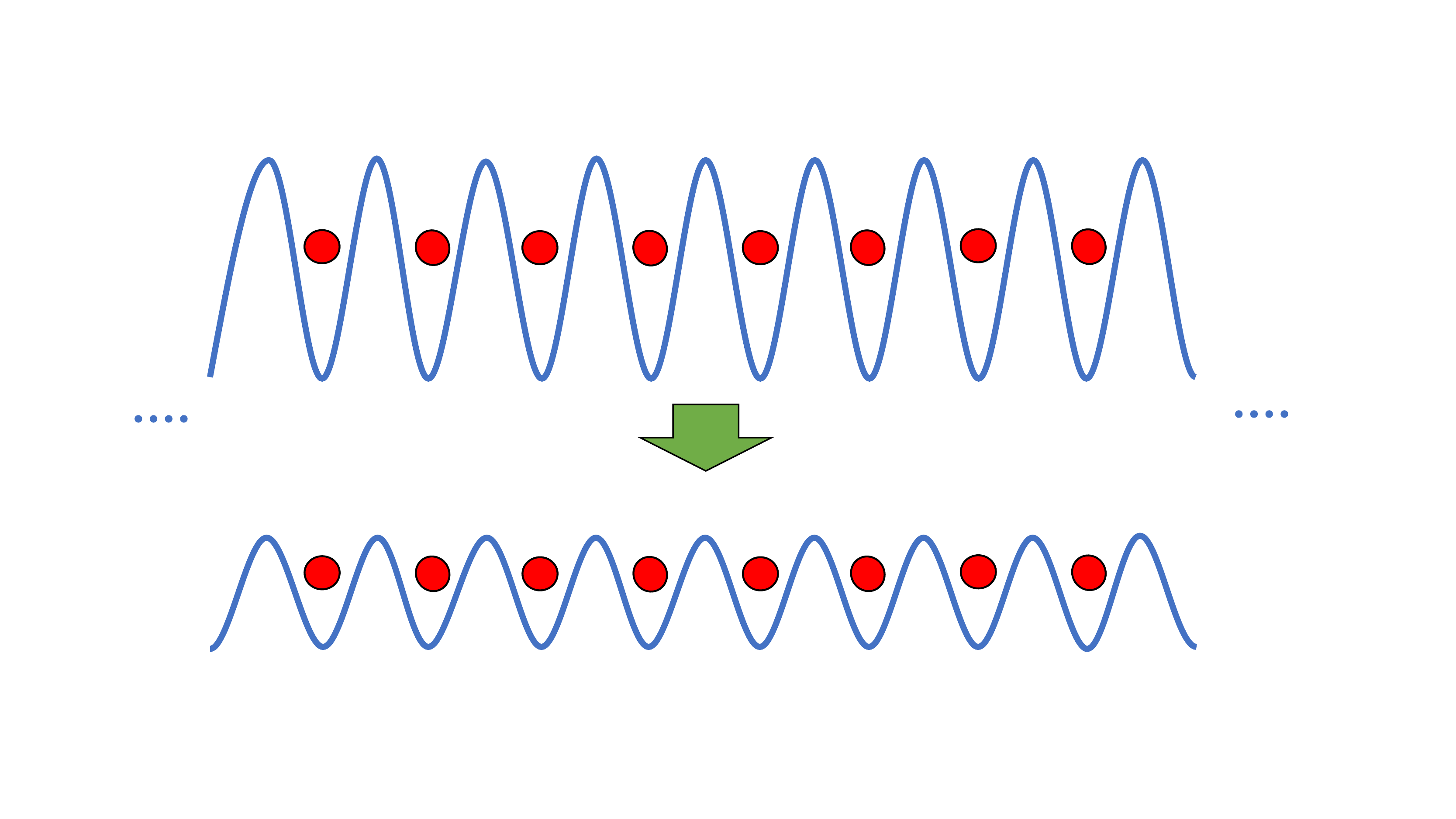}
\caption{We study the global non equilibrium behavior of two different optical lattice quenches. The ultracold atom gas is initially held in a deep optical lattice which is then either (a) removed or (b) merely  lowered. The post quench system is modelled by the Lieb-Liniger and Sine-Gordon Hamiltonians respectively. }\label{FigCartoon}
\end{figure}

We turn now to study  the global properties of the post quench system though the Loschmidt amplitude \eqref{LA} and work distribution function \eqref{P} focusing on the experimentally relevant case of a cold atom gas initially held in a deep optical lattice which is then either lowered or removed entirely  see FIG. \ref{FigCartoon}. In the first case the bosons evolve in a periodic potential and are described by the Sine Gordon model, in the latter case the system is translationally invariant and described by the Lieb Liniger model.

\subsection{The Gapless scenario - the Lieb Liniger bosons}

\begin{figure}
 \includegraphics[width=.45\textwidth]{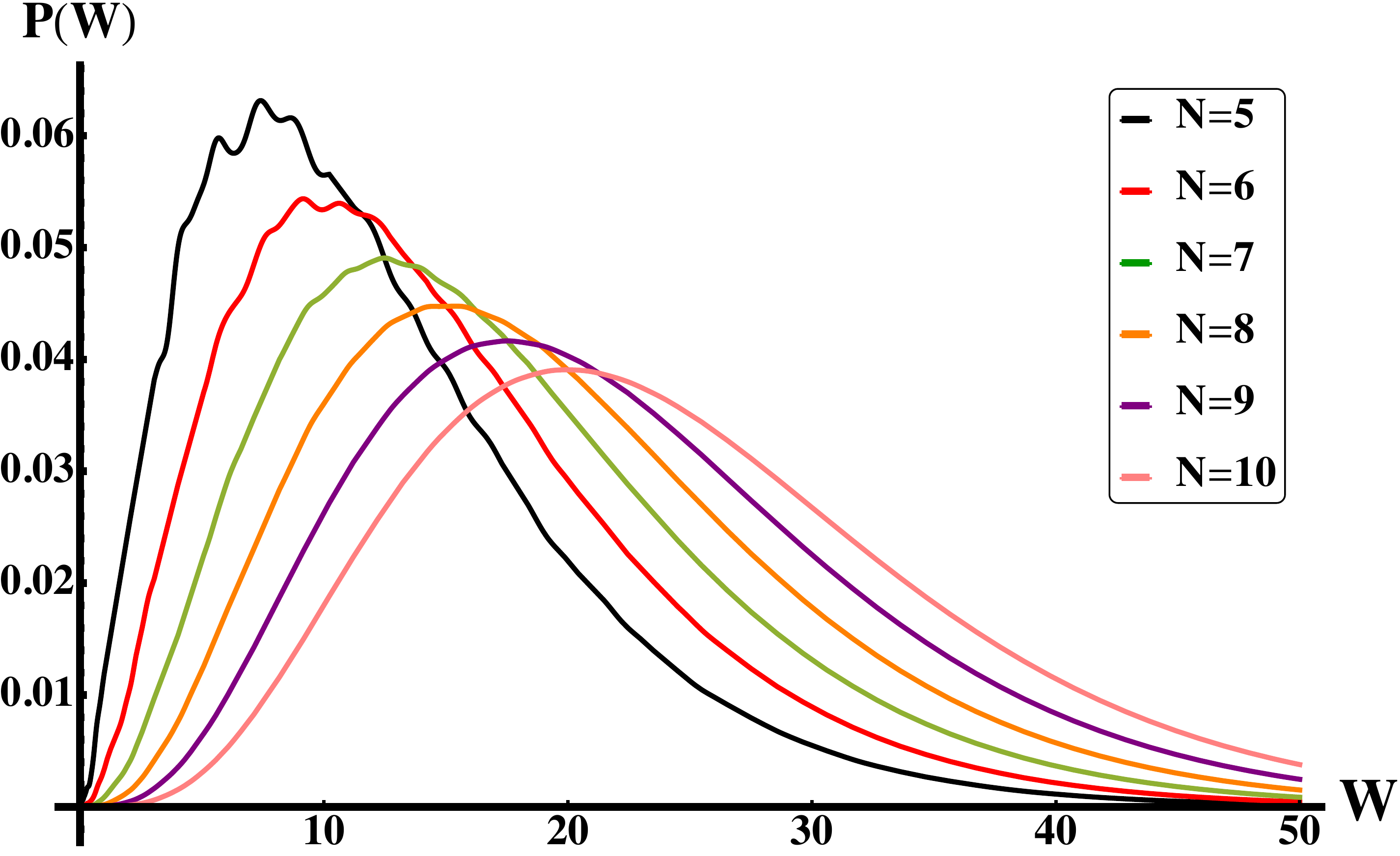}
  \includegraphics[width=.45\textwidth]{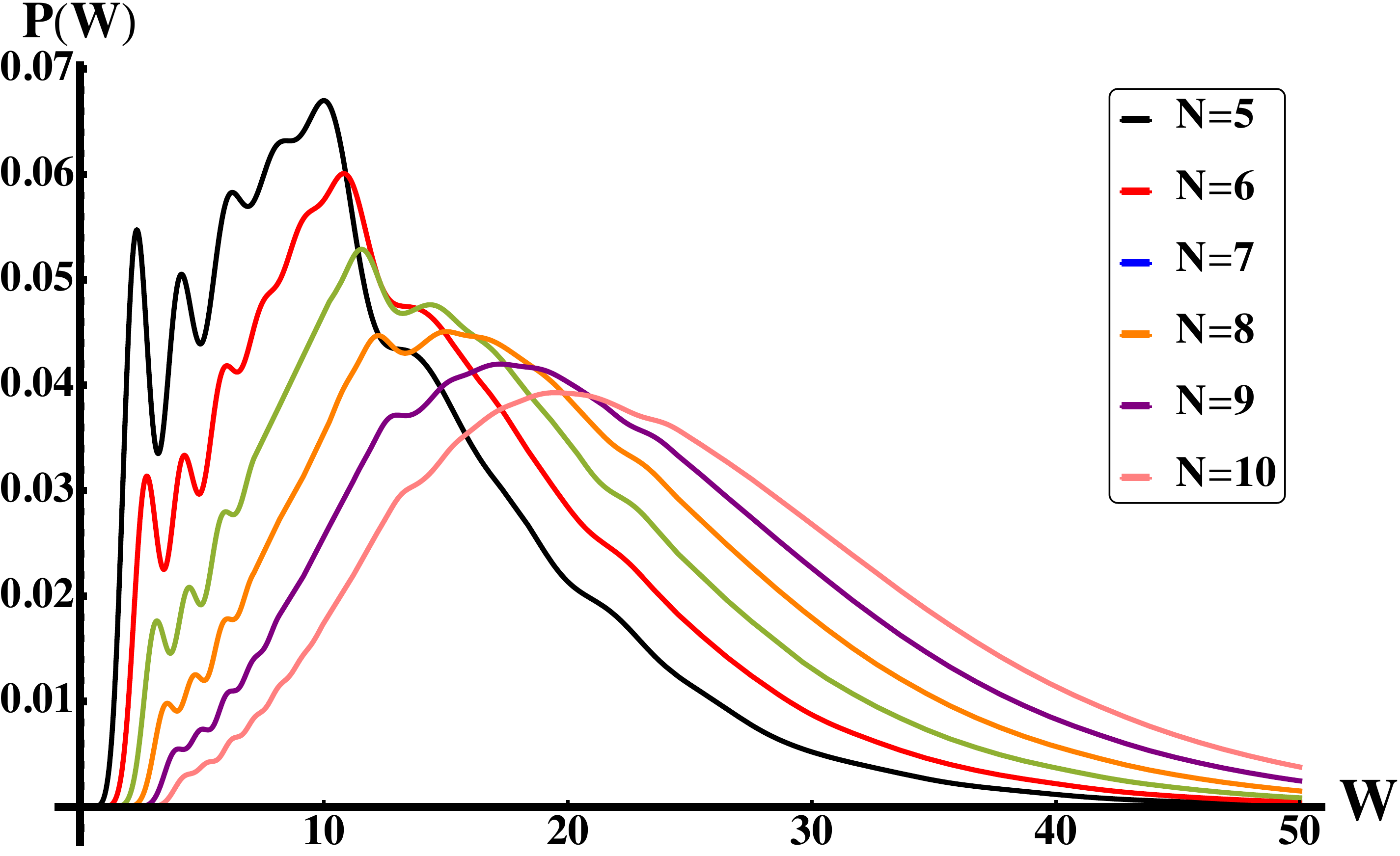}
\caption{The work distribution function, $\mathcal{P}(W)$, for different numbers of bosons released from an optical lattice  with $\delta/m=2$ and $\omega=10$. We measure the work from $\epsilon_i$ the initial state energy. On the left we show the distribution for non interacting bosons while on the right we show the same quantity for interacting bosons, $c>0$. Figure taken from \cite{RylandsAndreiLLwork}.   }\label{FigWork}
\end{figure}

To begin we consider the latter scenario where the system is initially described by the state \eqref{PSI0}  assuming that $N$ consecutive sites are filled. Furthermore  the unfilled part of the lattice is taken to be much larger than the filled portion so as to avoid complications arising from the boundary conditions. 
Employing the Yudson resolution of the identity, the Loschmidt amplitude can be determined to be \cite{RylandsAndreiLLwork},
\begin{eqnarray}\label{Gexact}
\mathcal{G}(t)&=&
\left[\frac{4\pi}{m\omega}\right]^{\frac{N}{2}}\!\!\sum_{n_1,\dots,n_N}e^{-\frac{1}{m\omega}\left[1+i\frac{\omega }{2}t\right]\sum_{j=1}^N\lambda_j^2}\frac{G(\{n\})}{\mathcal{N}(\{n\})}
\end{eqnarray}
where $G(\{n\})=\det{\left[e^{-i\lambda_j(\bar{x}_j-\bar{x}_k)-i\theta(j-k)\varphi(\lambda_j-\lambda_k)}\right]}$ and $\theta(j-k)$ is a Heaviside function. Using the same $1/c\delta$ expansion as before the Fourier transform of this can be explicitly found and analytic expressions for the work distribution, $\mathcal{P}(W)$ obtained. We plot this for both non interacting and strongly interacting bosons $c\delta\gg1$ in FIG. \ref{FigWork} for different particle number and see some commonalities as well as striking differences. Notice that the average work in both cases is the same, $\left<W\right>=N\omega/4$ as is the large $W>\left<W\right>$  behavior. The former statement can be understood from the fact that bosons are initially in non overlapping wave functions and $\left<W\right>=\matrixel{\Psi_0}{H}{\Psi_0}$. In comparison, the small $W\ll \left<W\right>$ behavior is strongly affected by the presence of interactions. Large resonant peaks are present in the interacting work distribution and  can be attributed to the scattering of strongly repulsive excitations in the post quench system. Those peaks which are closest to $\left<W\right>$ involve fewer scattering events while those $W=0$ involve more. As the particle number is increased these fluctuations are suppressed like $1/\sqrt{N}$ \cite{SotGamSil, Smacchia}.
For large systems of bosons the most interesting behavior therefore occurs in the region of $W\sim 0$ where the effects of the interaction are most keenly felt. In this region it can be shown that the distribution decays as a power law with the exponent drastically differing between the free and interacting cases. For the former we have $\mathcal{P}_{c=0}(W)\sim W^{\frac{N}{2}-1}$ whereas in the latter it is $\mathcal{P}_{c>0}(W)\sim W^{\frac{N^2}{2}-1}$, the presence of interactions in the system causing a dramatically faster decay of the work distribution. Behavior such as this will be seen in the next section also when the excitations are gapped as well as interacting.  

We can use our knowledge of $\mathcal{P}(W)$ to investigate the global behavior of the post quench system. As a consequence of the large $W$ agreement between the distributions for the interacting and non-interacting  systems we can determine that at short times $|\mathcal{G}(t)|^2$ is independent of the interactions. This corresponds to the initial period  of expansion  from the lattice in which the particles do not encounter one another. On the other hand,   small $W$ behavior
provides insight to the long time dynamics, the power law decay of $\mathcal{P}(W)$ near the origin translating to the long time power law decay of the LE. Fourier transforming the distribution for free bosons we find that as $t\to \infty$,  $|\mathcal{G}(t)|^2\to 1/t^N$ while in the interacting case  we have instead   $|\mathcal{G}(t)|^2\to 1/t^{N^2}$, a much faster decay. 
We attribute this dramatic difference in the decay away from the initial state to the fact that the large repulsive interactions acting on each other  forcing them to spread out into the one dimensional trap, thereby decreasing their overlap with $\ket{\Psi_i}$.
We should note that this is true regardless of the strength of the interactions and highlights the strongly coupled nature of even weakly interacting systems in low dimensions. As we saw earlier, in the long time limit any repulsive coupling flows in time strong coupling, therefore  the exponent is independent of the initial strength of $c$,  in the TG limit ($c=\infty$) one finds the same power law behavior at long times as for the finite $c$ case. This is the dynamical fermionization discussed in the previous section.

This analysis may be extended to the attractive regime of the Lieb Liniger model. Using \eqref{Identityatt} we find that up to additional bound state contributions we may continue the result for positive $c$ to negative values. Once again this is indicative of super-TG like behavior, this time in the global post quench behavior of the system. The bound state contributions are more evident when calculating $\mathcal{P}(W)$, even for large values of $|c|\delta$ and provide clearly identifiable signatures as opposed to the local properties discussed above. As was the case then they are exponentially suppressed and arise from transitions of the initial state to eigenstates containing bound states. Despite this suppression the bound states are clearly identifiable within $\mathcal{P}(W)$ due to the fact that they lower the energy and as a consequence there is a non vanishing probability of measuring $W<0$.  Importantly this does not violate the  2nd law of thermodynamics as the average work remains positive \cite{Jarzynski1} and coincides with the repulsive and free models,  $\left<W\right>=N\omega/4$. 

\subsection{The Gapped scenario - the Sine-Gordon bosons}
 \begin{figure}
  \includegraphics[width=.8\textwidth]{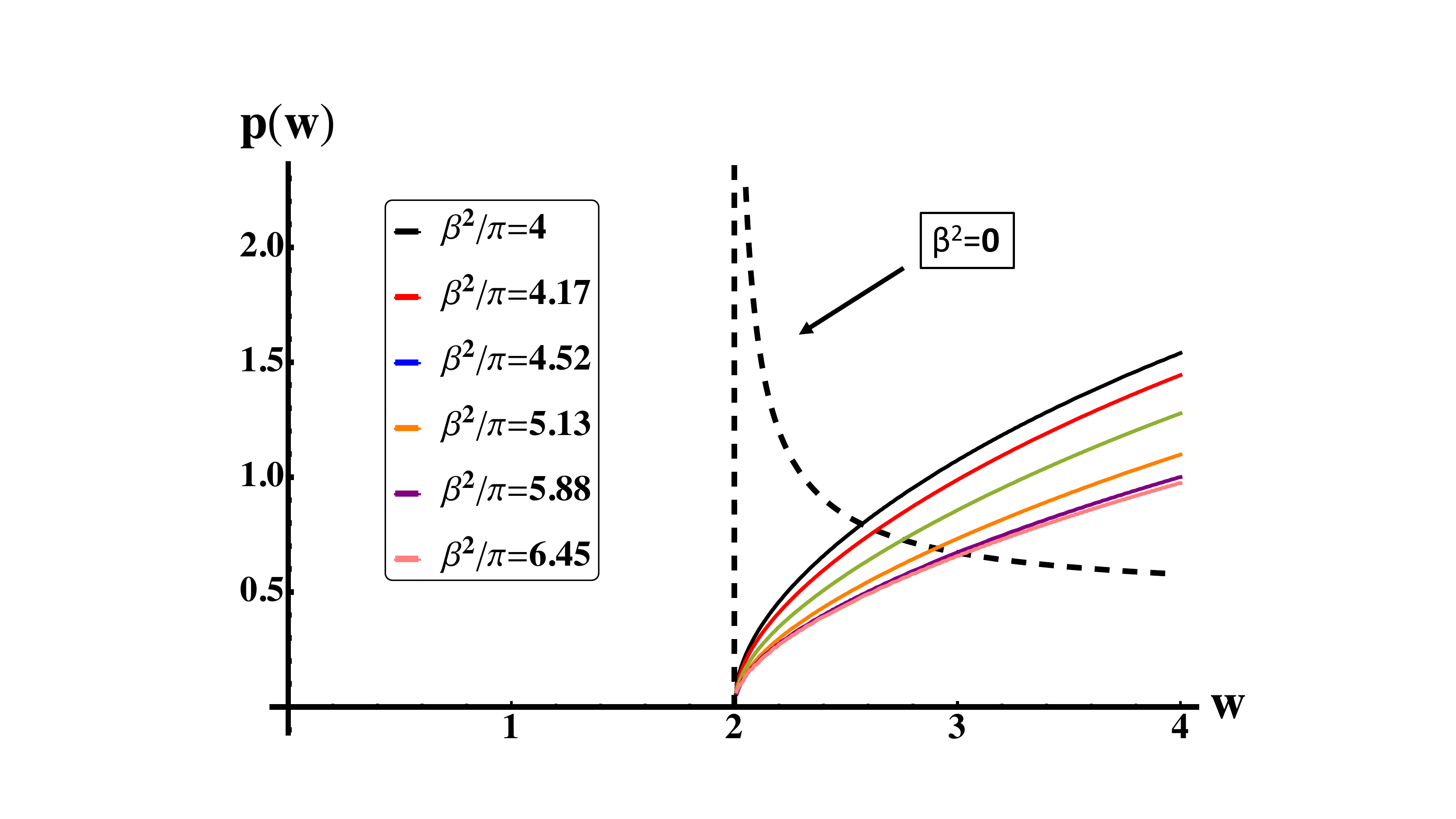}
\caption{The edge of the rescaled work distribution function, $p(w)=4\pi\mathcal{P}(W)/(m_\text{R}\mathcal{F}L)$ for quenches in the Sine-Gordon model. We measure work in units of the mass and from the ground state energy $w=(W-\epsilon_i+E_0)/m_\text{R}$. The dashed black line shows the same quantity for non interacting bosons, whilst the solid lines are different values of the interaction strength. The soild black line is equivalent to a quench for free fermions.  Figure adapted from \cite{RylandsMTM} }\label{FigMTMWork}
\end{figure}
We consider again a quench where a cold atomic gas is initially held in a deep optical lattice  which is  now only reduced rather than removed, see FIG. \ref{FigCartoon}.
The bosons of post quench system move in a periodic potential and the excitations develop a mass gap. The work distribution  has some  differences from the previous case and the effect of interactions on the system is more dramatic.  The system is  described by the Sine-Gordon (SG)  model \cite{FishFish, CazalillaRMP, TG}, given by 
\begin{eqnarray}
H_{\text{SG}}=\frac{v_s}{2\pi}\int \mathrm{d}x\,\Bigg\{\left[\nabla\phi(x)\right]^2+\left[\nabla\theta(x)\right]^2-M^2\cos{\left(\beta\phi(x)\right)}\Bigg\}
\end{eqnarray}
where $\phi(x), \nabla \theta(x)$ are canonically conjugate bosonic fields, $M$ is the strength of the periodic potential in which the bosons move, $\beta$ encodes the interaction and $v_s$ is the speed of sound in the system. We shall also compare the results to the bosonic non interacting case $\beta=0$ as well as  $\beta^2=4\pi$ which describes free fermions \cite{sGMTM, LutherEmery}. The Sine-Gordon model is ubiquitous in low dimensional physics and has many experimental realizations \cite{TG}.

  In what follows we will consider the so called irreversible work \cite{Esposito}  which measures the work done relative to the equivalent adiabatic process. Were we to lower the lattice in a reversible adiabatic fashion the work would exhibit no fluctuations and be given by  $W_\text{rev}=E_0-\epsilon_i$. Subtracting this value from the measured work we obtain the non trivial part of the work distribution function which comes from irreversible processes. Furthermore we shall focus on the region near $W\sim 0$ which in the thermodynamic limit displays universal characteristics \cite{SotGamSil}.

We set the scene using the non interacting case as an example. The $\beta=0$ limit is obtained replacing $-M^2\cos{(\beta\phi(x))}\to m^2\phi^2(x)$ leading to a quadratic Hamiltonian. The newly introduced $m$ is the mass of the bosonic excitations and is related to the lattice depth. The quench in this free boson system consist of taking $m_i\to m$ with $m_i\gg m$ and  can be solved by means of Bogoliubov rotation \cite{IucciCazalilla, CazalillaJstat}. In this fashion  it was shown that \cite{SotGamSil}
\begin{eqnarray}\label{Pfreebos}
\mathcal{P}(W)=\mathcal{F}\delta(W)+\theta(W-2m)\frac{\mathcal{F}LW}{\sqrt{W^2-4m^2}}+\dots
\end{eqnarray}
where $\mathcal{F}=|\braket{0}{\Psi_0}|^2$ is the overlap between the ground state and the initial state known as the fidelity and the ellipsis represent the part of $\mathcal{P}(W)$ which vanishes for $W<4m$.  Several distinct features have emerged in the presence of a gap. There is a delta function at $W=0$ which is weighted by the probability of the initial state transitioning to the ground state. Further the distribution vanishes for $0<W<2m$ and exhibits  power law behavior at threshold. As $W\to 2m$ from above, $P(W)\sim (W-2m)^{\alpha} $ with $\alpha=-1/2$ , similar to the X-ray edge phenomenon. This diverging edge exponent arises from the  initial state transitioning to a two particle excited state.  

As with the Lieb Liniger model these leading terms govern the long time dynamics. The presence of a gap and a sharp transition from the initial state to the ground state means that the LE decays to a constant value at long time, $|\mathcal{G}(t)|^2\to\mathcal{F}^2$,  this constant itself decays exponentially in system size. The approach to this value is governed by the exponent of the edge singularity which can be determined by Fourier transforming \eqref{Pfreebos} giving us  that as $t\to\infty$,  $|\mathcal{G}(t)|^2\to\mathcal{F}^2+\mathcal{O}(1/t^2)$. The much slower decay compared to the Lieb-Liniger model occurring as the long time behavior is dominated by  two particle excitations only rather than a macroscopic number. 

When interactions are present the Loschmidt amplitude is trickier to calculate. The spectrum of the model is entirely different, consisting of solitons and anti solitons which cannot be adiabatically connected to the free bosonic excitations of the $\beta=0$ model. The mass of the solitons is strongly renormalized to $m_\text{R}$ which depends upon the interaction strength and lattice depth, $M$ \cite{BT,ZamZam}.  Moreover,  for $0<\beta^2<4\pi$ the interactions between solitons and anti-solitons are attractive and bound states known as breathers may be formed. Using a standard trick one can map the LA to the partition function of a 2 dimensional classical model \cite{DDV1, CalabreseCardy2,AndSirk,  Palmai, poz1} which can then be computed exactly \cite{ poz2, poz3, Intquench, Baxter}. For the quench $M_i\to M$ with $M_i\gg M$, in the repulsive regime $\beta^2>4\pi$ one finds that in the region of $W\sim 2m_\text{R}$ the work distribution has the form \cite{RylandsMTM},
\begin{eqnarray}
\mathcal{P}(W)=\mathcal{F}\delta(W)+a\,\theta(W-2m_\text{R})\mathcal{F}\,L\,W\,\sqrt{\frac{W-2m_\text{R}}{W+2m_\text{R}}}+\dots
\end{eqnarray}
with $a$ a constant depending on $\beta$ and $M$. We see that as in the free case there is a delta function peak at the origin  followed by a gap up to $W=2m_\text{R}$ at which point  an X-ray edge like singularity  $\alpha=1/2$  appears. Therefore in contrast to the non interacting case the distribution function vanishes with a square root singularity at the threshold rather than diverging. This dramatic change in exponent results from a different pole structure in the overlap between the initial state and the state consisting of two excited quasi particles \cite{Palmai2}. FIG. \ref{FigMTMWork} shows
$\mathcal{P}(W)$ for several values of  $\beta^2$ as well as for the non interacting case $\beta=0$. The consequence for the system dynamics is that the echo approaches a constant value with a different exponent $|\mathcal{G}(t)|^2\to\mathcal{F}^2+\mathcal{O}(1/t^3)$. 

Once again the analysis can be extended to the attractive regime of the SG model where bound states are supported. The presence of bound states in the spectrum manifests itself through the appearance of additional delta functions in the region below the edge singularity,  $\sum_{n}\mathcal{F}_n\delta(W-m_{\text{B},n})$
where $m_{\text{B},n}$ are the masses of the bound states $|\text{B}_n\rangle$, $\mathcal{F}_n= \langle \text{B}_n |\phi_i\rangle$  is their overlap with the initial state and the sum is over even parity bound states only. These represent the transition to excited states consisting of single zero momentum bound state. Since these bound state excitations  are exact eigenstates of the model they have infinite lifetime and produce delta function in $\mathcal{P}(W)$ just as for the transition to the ground state.   The consequence for the dynamics of the system is that the at long time the LE  includes oscillatory contributions, for example  a single bound state gives $|\mathcal{G}(t)|^2\to \mathcal{F}^2+\mathcal{F}\mathcal{F}_1\cos{(m_{\text{B},1}t)}+\mathcal{O}(1/t^3)$. If integrability is weakly broken in the post quench Hamiltonian these  features will broaden due to the finite bound state lifetime and the long time oscillations will be damped.  

As pointed out before the measurement of work distribution can be systematically carried out in cold atoms experiment and provide detailed information on the underlying dynamics of the system.

\section{Summary and Outlook}
 In this review we have explored some aspects of the far from equilibrium behavior of integrable models. After a broad overview of the current status of the field we investigated some particular phenomena through a number of  illustrative examples. We saw that the Bethe Ansatz solution of the Lieb-Liniger, Heisenberg and Sine-Gordon models provided us with a powerful tool with which to study both the local and global, non-equilibrium behavior of these strongly coupled systems. The quench dynamics of more complex models such as the Gaudin-Yang model \cite{Gaudin, YangPRL67}  describing multi-component gases  has also been accessed via the Yudson approach  \cite{GuanAndrei}  allowing the study of  phenomena such a quantum Brownian motion or the dynamics of FFLO states \cite{FuldeFerrell, LarkinOvchinnikov}. Similarly the quench dynamics of other models such as the Kondo and Anderson models are currently  studied via  the Yudson approach \cite{CulverAndrei, TouraniAndrei}. They  give access to such quantities as  the time evolution of the Kondo resonance or  of the charge or heat currents in voltage or temperature driven two lead quantum dot system.

 These methods we discussed could be thought as being microscopic, starting from the exact eigenstates of the system. Recently these problems have been studied from a macroscopic perspective by combining integrability and ideas from hydrodynamics \cite{DoyonPRX, deNardisGHD}. Generalized hydrodynamics (GHD) provides a simple description of  the non equilibrium integrable models on long length scales and times. It has been utilized in studies of domain wall initial states in the Lieb-Linger and the emergence of light cones in quenches of the XXZ model \cite{BetheBoltz, Solvable}.  This method allows  the incorporation  integrability breaking effects within the formalism \cite{ANote}, but is limited to ``Euler scale" dynamics. It would be of great interest compare the results and expectations of GHD with the methods and results presented here to further understand the limitations of both the microscopic and macroscopic approaches.

\acknowledgements
We are grateful to Adrian Culver, Huijie Guan, Garry Goldstein, Deepak Iyer, Wenshuo Liu  and Roshan Tourani for many discussions.
This research was supported by DOE-BES (DESC0001911) (CR) an by NSF Grant DMR 1410583 (NA).

\bibliography{mybib}
\end{document}